\documentclass[aps, twocolumn, superscriptaddress, showpacs, nofootinbib, longbibliography]{revtex4-1}

\usepackage[utf8]{inputenc}
\usepackage[T1]{fontenc}
\usepackage{ae,aecompl} 
\usepackage{graphicx}
\usepackage{amsmath}
\usepackage{xcolor}
\usepackage{amssymb}
\usepackage{latexsym}
\usepackage{wasysym}
\usepackage{psfrag}
\usepackage{ifthen}
\usepackage[citecolor=blue,colorlinks=true]{hyperref}
\usepackage{float}
\usepackage[utf8]{inputenc}
\usepackage{lineno}
\usepackage{ulem}
\usepackage{multirow}
\usepackage{subcaption}
\usepackage{orcidlink}
\usepackage[nolist]{acronym}

\definecolor{blue-violet}{rgb}{0.54, 0.17, 0.89}

\begin{document}

\title{Leveraging cross-detector parameter consistency measures to enhance sensitivities of gravitational-wave searches} 

\author{Sayantan Ghosh \orcidlink{0009-0002-2884-6836}}
\affiliation{Department of Physics, Indian Institute of Technology Bombay, Mumbai, Maharashtra 400076, India}
\author{Leigh Smith \orcidlink{0000-0002-3035-0947}}
\affiliation{Dipartimento di Fisica, Università di Trieste, I-34127 Trieste, Italy}
\affiliation{INFN, Sezione di Trieste, I-34127 Trieste, Italy}
\author{Jiyoon Sun \orcidlink{0009-0008-8278-0077}}
\affiliation{National Institute for Mathematical Sciences, Daejeon 34047, Republic of Korea}
\affiliation{Department of Physics, Chung-Ang University, Seoul 06974, Republic of Korea}
\author{Archana Pai \orcidlink{0000-0003-3476-4589}}
\affiliation{Department of Physics, Indian Institute of Technology Bombay, Mumbai, Maharashtra 400076, India}
\author{Ik Siong Heng \orcidlink{0000-0002-1977-0019}}
\affiliation{SUPA, School of Physics and Astronomy, University of Glasgow, Glasgow G12 8QQ, United Kingdom}
\author{V. Gayathri \orcidlink{0000-0002-7167-9888}}
\affiliation{Leonard E. Parker Center for Gravitation, Cosmology, and Astrophysics, University of Wisconsin–Milwaukee, Milwaukee, WI 53201, USA}

\begin{abstract}

All-sky searches for generic short-duration astrophysical \ac{GW} signals are
often challenging because of noise transients. Developing novel signal-noise discriminators is crucial for
\ac{GW} transient searches with LIGO, Virgo, and KAGRA (LVK) detectors. In this work, we demonstrate the sensitivity improvement of a weakly-modeled \ac{GW} transient search by leveraging a recently developed Jensen Shannon divergence (JSD)-based cross-detector parameter consistency measure. We first extend a 2-detector JSD-based
measure, developed in an earlier work \cite{Ghosh_2023}, to a 3-detector network. We leverage this to modify the
test statistic of the existing coherent WaveBurst (cWB)-Gaussian Mixture Modelling (GMM) algorithm
for short-duration transients towards improving the search sensitivity to ad-hoc waveforms like Sine-Gaussians, Gaussian Pulses, and White Noise Bursts. We find that with the new method, which we term cWB-GMM-JSD, the sensitivity to the ad-hoc waveforms, given by $h_{\mathrm{rss50}}$ (the signal amplitude at which 50$\%$ of the injected signals can be detected), improves by $\sim 10-20 \%$ at an \ac{IFAR} of 10 years for the 2-detector network consisting of \ac{LHO} and \ac{LLO} detectors, and by $\sim 5-10 \%$ at the same \ac{IFAR} for the 3-detector network consisting of \ac{LHO}, \ac{LLO} and Virgo detectors. Finally, we apply the modified statistic in the revised data analysis pipeline on the publicly available data from the third observing run (O3) of the LIGO and Virgo detectors. Although we do not find any new event in the O3 data, we see a notable rise in the statistical significance of
most of the known \ac{GW} events, which further testifies to the enhancement in sensitivities.

\end{abstract}

\maketitle

\acrodef{ROC}[ROC]{Receiver Operating Characteristic}
\acrodef{LR}[LR]{likelihood ratio}
\acrodef{GWOSC}[GWOSC]{Gravitational Wave Open Science Center}
\acrodef{KDE}[KDE]{kernel density estimate}
\acrodefplural{KDE}[KDEs]{kernel density estimates}
\acrodef{BBH}[BBH]{binary black hole}
\acrodef{BNS}[BNS]{binary neutron star}
\acrodef{NSBH}[NSBH]{neutron star black hole}
\acrodef{BH}[BH]{black hole}
\acrodef{LVC}[LVC]{LIGO Scientific and Virgo Collaborations}
\acrodef{GW}[GW]{gravitational-wave}
\acrodefplural{GW}[GWs]{gravitational waves}
\acrodef{CBC}[CBC]{compact binary coalescence}
\acrodefplural{CBC}[CBCs]{compact binary coalescences}
\acrodef{CI}[CI]{confidence interval}
\acrodef{IMBH}[IMBH]{intermediate-mass black hole}
\acrodef{SMBH}[SMBH]{supermassive black hole}
\acrodefplural{SMBH}[SMBHs]{supermassive black holes}
\acrodefplural{IMBH}[IMBHs]{intermediate-mass black holes}
\acrodef{SNR}[SNR]{signal-to-noise ratio}
\acrodef{FAR}[FAR]{false alarm rate}
\acrodef{PSD}[PSD]{power spectral density}
\acrodefplural{PSD}[PSDs]{power spectral densities}
\acrodef{LVK}[LVK]{LIGO, Virgo and KAGRA}
\acrodef{GR}[GR]{General Relativity}
\acrodef{FF}[FF]{fitting factor}
\acrodef{O3}[O3]{third observing run} \acrodef{GWTC}[GWTC]{Gravitational Wave Transient Catalogue}  
\acrodefplural{GWTC}[GWTCs]{Gravitational Wave Transient Catalogues}
\acrodef{IFAR}[IFAR]{inverse false alarm rate}
\acrodefplural{IFAR}[IFARs]{inverse False Alarm Rates}
\acrodef{BHB}[BHB]{black hole binary}
\acrodefplural{BHB}[BHBs]{black hole binaries}
\acrodef{LHO}[LHO]{LIGO-Hanford}
\acrodef{LLO}[LLO]{LIGO-Livingston}
\acrodef{JSD}[JSD]{Jensen Shannon divergence}
\acrodef{IGWN}[IGWN]{International Gravitational-Wave Observatory Network}
\acrodef{cWB}[cWB]{coherent WaveBurst}
\acrodef{PE}[PE]{parameter estimation}
\acrodef{WDM}[WDM]{Wilson-Daubechiers-Meyer}
\acrodef{GMM}[GMM]{Gaussian Mixture Modeling}
\acrodef{EM}[EM]{Expectation Maximization}
\acrodef{WNBs}[WNBs]{White Noise Bursts}
\acrodef{SGs}[SGs]{Sine Gaussians}
\acrodef{GAs}[GAs]{Gaussian Pulses}
\acrodef{CCSN}[CCSN]{core collapse supernovae}
\acrodef{ML}[ML]{machine learning}
\acrodef{XGB}[XGB]{XGBoost}
\acrodef{GMM}[GMM]{Gaussian Mixture Modelling}

\section{Introduction}
\label{Sec:intro}

The detection of the \acf{GW} event GW150914 on 14th September, 2015 \cite{LIGOScientific:GW150914}, ushered in a new era of observational astronomy. Since then, the ground-based \ac{GW} detectors, operated by the \ac{LVK} collaboration, have observed around 100 \ac{GW} events over three observing runs - O1, O2 and O3 \cite{GWTC-1,GWTC-2,GWTC-2.1,GWTC-3} and around 200 events in O4 so far. All of these signals have been consistent with \acp{CBC} involving black holes and neutron stars. A vast majority of these events have been declared to be \ac{BBH} mergers, that is, mergers of stellar-mass black holes. The remaining few include \ac{BNS} systems \cite{gw170817-discovery,gw170817-properties,gw190425-discovery}, \ac{NSBH} systems \cite{2nsbh-discoveries}, and the first confident \ac{IMBH} binary \cite{gw190521-discovery,gw190521-properties,o3-imbh}. However, the current ground-based \ac{GW} detectors are capable of detecting more than just \acp{CBC}. They are sensitive to \acp{GW} emitted by core-collapse supernovae (CCSN) \cite{Powell:2018-s18,Powell:2020-m39,Radice:2018-s9,OConnor:2018-m20,Abdikamalov:2020jzn,CCSN-o3-Szczepanczyk}, cosmic strings \cite{CosmicString-o1-LVK,CosmicString-o3-LVK}, hyperbolic black hole encounters \cite{Cho:2018hyp,Morras:2021hypgw,Dandapat:2023hyp,Bini:2023hyp}, radiation driven capture \cite{Bae:2017capture,Ebersold:2022cap}, non-linear memory effects \cite{NL-memory-ebersold20,Hubner:2019NL-mem}, and neutron star glitches \cite{NSgkitch-2006-LVK,Lopez:2022NS-glitch,Yim:2020NS-glitch}. The waveforms of many of these astrophysical sources are either not well-known or are computationally too expensive for use in matched filtering searches \cite{LIGOScientific:matchedfilter}, which perform well for well-modeled \ac{CBC} signals. Searches for generic \ac{GW} transients or \textit{bursts} adopt weakly-modeled approaches which capture generic signal morphologies. Often \ac{GW} searches are categorized based on the duration of the target signals and tuned separately. The \textit{all-sky short} search conducted by the \ac{LVK} collaboration searches for signals with duration of the order of a few seconds or less with little to no assumptions on the nature of the signal \cite{allskyo3}. While the weakly-modeled nature of the search does make it sensitive to a wide variety of signal morphologies, it also makes the search vulnerable to short-duration non-stationary, non-Gaussian noise transients or \textit{glitches}, some classes of which are morphologically similar to \ac{GW} signals.  

With improvements in  detector sensitivities with every observation run, the \ac{GW} data is often plagued with new types of noise transients of terrestrial origin. The development of data analysis techniques for identifying glitches and mitigating their effect on \ac{GW} searches is one of the major thrust areas in \ac{GW} research. Glitch-removal can be carried out in different stages. There are different types of vetoes \cite{vetoes} and gating \cite{pycbc-paper} which remove segments of data contaminated by non-stationary behavior before running a search pipeline on the data. There are specially designed test statistics \cite{chisquared} that can discriminate between signals and glitches. The coherence test developed in \cite{veitch} and BayesWave \cite{bayeswave,bayeswave-glitch-subtraction} approach this problem in Bayesian ways. The former computes the Bayes Factor between the hypothesis that the data contains a coherent \ac{CBC} signal in the detector network ($H_{\mathrm{coh}}$) against the hypothesis that the data contains incoherent glitches ($H_{\mathrm{inc}}$). BayesWave builds models for both signals and glitches by projecting both signals and glitches onto the Morlet-Gabor wavelet basis. However, many glitches still remain in the data even after the application of vetoes, gating and other glitch-removal techniques.

One of the leading data analysis burst pipelines used to search for short-duration bursts in \ac{LVK} observing runs \cite{allsky-o1,allsky-o2,allskyo3} is \ac{cWB} \cite{Klimenko_2004wavelet,Klimenko_2008cwb,Klimenko:2015cwb,Drago:2020cwb} which has contributed significantly to the \ac{GW} transient catalogs \cite{GWTC-1,GWTC-2,GWTC-3}. \ac{cWB} searches for \acp{GW} by calculating the excess coherent power within a network of \ac{GW} detectors, making minimal assumptions on the signal morphology. It calculates a set of attributes which quantify the degree of coherence of the signal across the detector network. These attributes follow different distributions for signals and noise. Earlier, cuts were placed manually on the various attributes, but recently, \ac{ML} approaches are being utilized to post-process the \ac{cWB} attributes, and this improves our ability to distinguish \ac{GW} signals from glitches. 

The two \ac{ML} post-processing algorithms that are employed currently are eXtreme-Gradient Boost (or \ac{XGB}) and \ac{GMM}. \ac{XGB} \cite{xgboost_2021,xgboost_2023} uses a decision tree-based ensemble learning classifying algorithm to construct a penalty factor which is multiplied with the \ac{cWB} ranking statistic, widening the separation between the signal and noise distributions.
\ac{GMM} models the distribution of multi-dimensional \ac{cWB} attributes as a superposition of Gaussians. The signal and noise populations are modeled separately and a likelihood ratio statistic is used to assess the detection significance. This approach to post-process the \ac{cWB} attributes was introduced in \cite{Gayathri_2020}, and its upgraded versions were described in \cite{Lopez_2022,Smith_2024}. Here, we build on the latest version detailed in \cite{Smith_2024}, which we refer to as cWB-GMM throughout this article.

In \cite{Ghosh_2023}, the authors proposed an independent \ac{JSD}-based metric which quantifies the consistency of parameter estimates across two detectors to distinguish between massive black hole binary signals and coincident glitches. In this work, we demonstrate that this metric can be adapted to modify the detection statistic in a way that increases the detection sensitivity. The \ac{JSD}-based parameter consistency measure was introduced in \cite{Ghosh_2023} for a 2-detector network. In this work, we extend the method to a 3-detector network, taking into account differences in their sensitivities. With the generic ad-hoc signals, we show that the metric can be integrated with the existing burst pipelines leading to a substantial improvement in the detection sensitivity. In this work, we demonstrate the application of the \ac{JSD} method on the cWB-GMM pipeline. We find that this improves the sensitivity of the pipeline (quantified by $h_{rss50}$ - root sum squared strain at $50\%$ detection efficiency), to various ad-hoc waveforms like \ac{GAs}, \ac{SGs} and \ac{WNBs}, which are used for testing the performance of weakly-modeled searches. The percentage improvement depends on the parameters characterizing these ad-hoc waveforms and the detector network, but on the whole, we see about $10-20 \%$ improvement for the 2-detector \ac{LHO}-\ac{LLO} (LH) network and about $5-10 \%$ improvement for the 3-detector \ac{LHO}-\ac{LLO}-Virgo (LHV) network. In addition to the ad-hoc waveforms we apply our method to several \ac{CCSN} simulations, for which we see an improvement of $\sim 5-10 \%$ for both the LH and LHV networks at an \ac{IFAR} of 10 years. The only \ac{CCSN} simulation which shows a serious degradation in the sensitivity is the magneto-rotational waveform \texttt{Abdikamalov\_A4O01.0}, and this is seen only for the LHV network.

This work is organized as follows. Section \ref{Sec:Methodology} first discusses the extension of the 2-detector \ac{JSD} method introduced in an earlier work to a network with arbitrary number of detectors. It then outlines the current cWB-GMM pipeline and subsequently introduces the modified ranking statistic, which combines the cWB-GMM ranking statistic and the proposed \ac{JSD}-based metric. Section  \ref{Sec:data} describes the data used in this work. Section \ref{Sec:results} showcases the performance of the \ac{JSD}-based measure as a signal-noise discriminator and demonstrates the improvement in pipeline sensitivity due to the proposed modification. This section also presents the re-analysis of the O3 data from the LIGO-Virgo detectors using the updated methodology, and the work concludes with section \ref{Sec:conclusions}.

\section{Methodology}
\label{Sec:Methodology}

In this version of cWB-GMM, we use the \ac{JSD}-based multi-detector parameter consistency measure to revise the test statistic with a view to improving its sensitivity. The integration of this new parameter consistency measure with the existing cWB-GMM pipeline increases its ability to discriminate between noise transients and short-duration astrophysical transients.

\subsection{Extension of Jensen Shannon Divergence to an arbitrary number of detectors}
\label{jsd-2d-to-3d}

\ac{JSD} is a bounded and symmetric measure of the dissimilarity between two probability distributions \cite{jsd-ref}. In \cite{Ghosh_2023}, the authors borrowed this idea to define a \ac{JSD}-based measure of astrophysical parameter consistency between 2 detectors. Here, we adopt it to revise the test statistic so that the detection efficiency of cWB-GMM can be improved.

We first remind the reader of the main results obtained in \cite{Ghosh_2023}. If $p_1(\theta)$ and $p_2(\theta)$ are the posteriors of the parameter $\theta$ in detectors 1 and 2, then 
\begin{equation} \label{jsd-defn}
\begin{split}
    \mathrm{JSD}_{\theta}(p_1(\theta)||p_2(\theta)) = \frac{1}{2}\int p_1(\theta)\log_2\frac{p_1(\theta)}{p_\mathrm{avg}(\theta)} d\theta + \\
    \frac{1}{2}\int p_2(\theta)\log_2\frac{p_2(\theta)}{p_\mathrm{avg}(\theta)} d\theta  \, ,
\end{split}
\end{equation}
where $p_{\mathrm{avg}} = \frac{1}{2}(p_1(\theta) + p_2(\theta))$. This quantity is close to 0 for consistent distributions, and closer to unity for inconsistent distributions. In \cite{Ghosh_2023}, it was shown that
$\mathrm{JSD}(p_H(\theta)||p_L(\theta))$ is small for \ac{IMBH} binary signals having detector-frame total mass in the range $(200,500) \mathrm{M}_{\odot}$ and high for coincident noise triggers which resemble \ac{IMBH} binaries morphologically.

In this work, we extend the approach outlined in \cite{Ghosh_2023}, to an arbitrary number of detectors with arbitrary noise sensitivities.
We introduce a way to combine $\mathrm{JSD}(p_1(\theta)||p_2(\theta))$ for different pairs of detectors, taking into account differences in their sensitivities and construct a single quantity $\mathcal{J}$ which can be interpreted as a measure of astrophysical parameter consistency across the detector network.

Consider an arbitrary (at least 2) number of ground-based \ac{GW} detectors. Let the detectors be labeled by indices $i,j,k...$ and let the parameters be labeled by the index $p$. Let $J_p^{ij}$ be the \ac{JSD} computed for the parameter $p$ between the detectors $i$ and $j$, following Eq. \ref{jsd-defn}. Let $\sigma_p^i$ be the standard deviation of the posterior of the parameter $p$ in the detector $i$. 
We define:
\begin{equation}
    W_p^{ij} = \frac{1}{\sigma_p^i \sigma_p^j}
\end{equation}

A posterior with smaller $\sigma_p^i$ is more localized and hence contains more information than a posterior having higher $\sigma_p^i$. Thus, $\frac{1}{\sigma_p^i}$ can be thought of as the relative importance of the information contained in the posterior for the $i$-th detector and $W_p^{ij}$ is a measure of the relative importance of the combination of detectors $i$ and $j$ (the most sensitive pair of detectors is the one with the highest value of $W_p^{ij}$).


We now define the parameter specific combined \ac{JSD} for a network of detectors as the sum of $J_p^{ij} $s weighted by $W_p^{ij}$ as:
\begin{equation}
    J_p = \frac{\sum_{i < j} J_p^{ij} W_p^{ij}}{\sum_{i < j} W_p^{ij}}
\end{equation}
where the summation over $i < j$ indicates summation over all possible detector pairs in the network, each pair considered only once. Finally, we define the multi-detector network JSD $\mathcal{J}$ for a given GW network by averaging over the total number of parameters $N_p$ as:
\begin{equation}\label{eq:J_val}
    \mathcal{J} = \frac{1}{N_p}\sum_p J_p
\end{equation}

In this work, we calculate $\mathcal{J}$ for a 2-detector network comprising \ac{LHO} and \ac{LLO} detectors, and a 3-detector network comprising \ac{LHO}, \ac{LLO} and Virgo detectors. The calculation of $\mathcal{J}$ requires us to run \ac{PE} with the strain data coming from the detectors. In general, the aim of PE is to obtain precise and accurate estimates of underlying astrophysical parameters of the signal. In this work, however, our purpose is not to obtain accurate parameter estimates, but rather
to check whether the parameter estimates in different detectors are consistent with one another. After all, PE is a projection of the data onto waveforms drawn from a prior defined in the parameter space. If waveforms
in a particular region of the parameter space describe the signal present in the data well, then the posterior peaks in that region, indicating that the projection of the data onto those waveforms is high. If the incoming signal is from a definite astrophysical source characterized by a definite morphology, we expect the parameters to be consistent across the detectors, though they
may not be accurate as the accuracy crucially depends on the completeness of the waveform model used. On the other hand, glitches occurring in different detectors are caused by different physical factors and thus have different morphologies. This means that the single-detector
parameter estimates of coincident glitches are expected to be inconsistent between detectors. Thus, as long as we use the same waveform model in the \ac{PE} run in all the detectors, there is a
possibility of using similarity measures between parameter estimates as signal-noise discriminators. This will be our usage of the \ac{PE} results in this work. \textit{This would be the first work where the PE-based consistency measure is combined with the detection statistic to improve the detection significance.}   

We demand that the designed search algorithm be ``blind'' to the nature of the transient in the data. 
However, as this is a search for generic transients, it is impossible to find a single astrophysical model which accurately describes all our target signals. At best, we can choose a waveform model which loosely approximates most of our target signal types and is also easy to compute.  It was seen that the loudest glitches limiting the search sensitivity of all-sky short searches were mostly Blip, Tomte and Koi Fish glitches. In \cite{Ghosh_2023}, we had observed that such glitches were mapped well to the \ac{CBC} parameter space and the estimates were inconsistent between detectors. Further, massive \ac{CBC} signals are of short duration, with few cycles, amplitude peaking at merger time and decaying in the ringdown phase - thereby loosely capturing the main features of our target signals. Based on these observations, we choose an aligned-spin, dominant-mode-only, frequency domain \ac{CBC} model - \texttt{IMRPhenomXAS} and test whether it indeed yields consistent estimates for the signal types considered in this work \footnote{We acknowledge that it may be possible to further optimize the choice of waveforms and improve the separation between signal and glitch populations. We leave such a study for future work.}. We consider only the intrinsic parameters in the chosen waveform model as they dictate the internal dynamics of the system and hence leave an imprint in the signal morphology in each detector. We choose the detector-frame component masses $m_1$ and $m_2$, and the effective spin parameter $\chi_{\mathrm{eff}}$ for calculating $\mathcal{J}$ using Eqn. \ref{eq:J_val}. We marginalize over the extrinsic parameters like the distance, sky location, coalescence phase and time.  

The \ac{PE} runs are performed using the \texttt{PyCBC-Inference} \cite{pycbc-inference-paper} package and the \texttt{Dynesty} sampler \cite{dynesty-paper}. Our component mass prior is uniform in the detector-frame component masses in the range $(5,600)\mathrm{M}_{\odot}$. The luminosity distance prior is a quadratic power law in the range $(100,5000)\mathrm{Mpc}.$ The spins are aligned and their priors are uniform between -0.9 and +0.9. The sky location prior is uniform in solid angle all over the sky. We take a lower frequency of 15Hz and 500 live points.

\subsection{Overview of cWB-GMM}
\label{cwb-gmm-overview}
Coherent WaveBurst (cWB) \cite{Klimenko_2004wavelet,Klimenko_2008cwb,Klimenko:2015cwb,Drago:2020cwb} is one of the most extensively used weakly-modeled search algorithms, which searches for \ac{GW} signals buried in detector noise by correlating excess coherent power across the detector network. It first converts the strain data in the time domain to a time-frequency representation via the \ac{WDM} transformation \cite{Necula_2012}, and then clusters pixels with coherent energy above a certain threshold using nearest-neighbor algorithms. Such clusters are labeled \textit{triggers} or possible signals. Subsequently, it calculates a set of attributes which are measures of energy correlation between detectors and veto parameters which aid in the identification and rejection of noise artefacts. These attributes follow different distributions for astrophysical signals and noisy glitches enabling \ac{cWB} to set thresholds on these statistics and thereby distinguish between astrophysical signals and noise glitches.

\ac{GMM} is a supervised \ac{ML} algorithm which models a multi-modal distribution of data points in a multi-dimensional space as a weighted sum of multivariate Gaussian distributions. Each of these Gaussian distributions represents a subpopulation of the data set and is characterized by a mean vector, a covariance matrix and a weight. The number of Gaussian components used to represent the entire data set is a hyperparameter that can be chosen based on requirements specific to the problem at hand. Once the number of Gaussians is specified, the means, covariances and weights of the different components are obtained by the \ac{EM} technique \cite{EM_1977}, which maximizes the total likelihood of all the points in the data set belonging to the mixture model under consideration. 

Applied to \ac{cWB} triggers as a post-production method, \ac{GMM} provides an elegant way to model distributions in the \ac{cWB} attribute space. Thus, training separate GMMs for signal and noise triggers enables the construction of a likelihood ratio statistic:   
\begin{equation}
    T = L_S - L_N \,
\end{equation}
where $L_S$ and $L_N$ are the log likelihoods of a trigger belonging to the signal and the noise distributions respectively. 

For a test trigger, the likelihood ratio determines whether it is more likely to be a signal or noise event. A higher $T$ value indicates a stronger signal-like nature, while a lower $T$ value suggests a noise-like trigger. This way of mapping every point in the multidimensional attribute space to a single log-likelihood ratio statistic enables systematic classification and eliminates the need for manually setting thresholds for various \ac{cWB} attributes. The application of \ac{GMM} in the post-production of \ac{cWB} attributes has been detailed in earlier works \cite{Lopez_2022,Gayathri_2020,Smith_2024} and in this work we will use the latest version of the pipeline as described in \cite{Smith_2024}.

\subsection{cWB-GMM-JSD: A revised version of cWB-GMM} 
\label{cwb-gmm-jsd-test-statistic}

It was observed that the sensitivity of the cWB-GMM pipeline was limited by the presence of loud noise triggers which mimic astrophysical signals despite being of instrumental or environmental origin. This motivates us to make further modifications to the pipeline with a view to enhancing the sensitivity. In this work, we propose a \ac{JSD}-based modification to the GMM log likelihood ratio used in \cite{Smith_2024}, which penalizes triggers which are of noise origin.
As seen in \cite{Ghosh_2023} and as shown later in Sec. \ref{cwb-gmm-jsd-in-action}, $\mathcal{J}$ takes low values for signals and high values for noise triggers. This motivates us to define the revised ranking statistic as:
\begin{equation}\label{TJ-defn}
    T_J = 
    \left\{ 
    \begin{array}{c l}
        \frac{T - T_0}{(1 + \mathcal{J})^n} + T_0, & \quad \text{if} \quad T \geq T_0 \\ 
        T, & \quad \text{otherwise}
    \end{array}
    \right.
\end{equation}

where $n$ is a positive integer and $T_0$ is the threshold of $T$ above which we make the proposed \ac{JSD}-based modification. This threshold is necessary since the computational cost associated with the above modification depends to a great extent on the speed of the \ac{PE} runs, and it may not be feasible to perform \ac{PE} for all the triggers. 

The rationale for choosing this particular form of modifying the detection statistic is the following: for signals, $\mathcal{J} \ll 1$ and so $T_J \approx T$, that is, the values of the detection statistic for signals remain largely unaltered. For noise triggers, $\mathcal{J} \sim 1$ and hence $T_J \sim T_0$, i.e., the background triggers which had $T \geq T_0$ will now be assigned a much lower value of $T_J$ close to $T_0$. In essence, most of the signals will retain their ranking, whereas the noise triggers with ranking statistic above $T_0$ will be down-ranked to values close to $T_0$. This will conduce to the detection of signals with greater significance, which were earlier obfuscated by the noise triggers above the threshold. 

\subsubsection{Choice of $T_0$}
The threshold choice is determined by a trade-off between the following 2 considerations:
\begin{enumerate}
    \item Ideally, we would like to reweigh all the triggers and so $T_0$ should be set to the lowest $T$ value in our dataset (which is typically a large negative number). 
    \item But as $T_0$ is lowered, the number of triggers also increases non-linearly, making the \ac{PE} of all the triggers computationally intensive and intractable.    
\end{enumerate}

In this proof-of-principle work, we take $T_0 = 0$. This choice is justifiable for 2 reasons: First, as $T$ is the log-likelihood ratio between the signal and noise models, candidates with negative values of this quantity lie in a region of the attribute space where the noise likelihood dominates over the signal likelihood, indicating that the candidates have a greater probability of being noise triggers. Second, for all the signal types considered in this work, it was observed that more than $95\%$ of the signals have $T \geq 0$. The choice of $T_0$ is based on simulations so that most of the triggers of noise origin are down-ranked.

\subsubsection{Choice of n}
We need to tune $n$ based upon some generic simulation set representative of the search parameter space and the noise statistics. The $\mathcal{J}$ distributions may vary with the noise \ac{PSD} of the detectors, the loudness of the injected signals, and the noise distribution, thus making the choice of $n$ dependent on the data.  
The value of $n$ is determined by a trade-off between the following 2 effects:
\begin{enumerate}
    \item A higher $n$ will always lead to a better reduction of the noise, which is always desirable.
    \item A higher $n$ will also run the risk of dragging down the signal $T_J$ values. In other words, if $n$ is too high, $T_J$ will start deviating from $T$, and that will counteract the effect of noise reduction. 
\end{enumerate}
In this work, we demonstrate the sensitivity improvement with $n=4$. Beyond this, we start seeing a significant drop in the $T_J$ values of signals having lower frequencies and longer duration among the suite of waveforms which we are considering.

\section{Sensitivity estimation with playground data}
\label{Sec:data}
The calibrated strain data provides the detector's response to the astrophysical sources, which is often dominated by detector noise stochastic in nature and greater than that of the underlying astrophysical signal. The search algorithm identifies time periods  at which potential astrophysical signals or \textit{triggers} are present and assigns statistical significance to each trigger. A trigger is flagged as a potential astrophysical candidate only if this statistical significance is above a pre-determined threshold.
 
Before we apply the search, we assess the detection sensitivity of the pipeline utilizing the playground data with target signals injected in the detector noise. The complete sensitivity study involves the detailed summary of the playground data, distribution of the noise background for assigning statistical significance to the injected signals and the metric for sensitivity computed for the injected signals. This section will summarize all the key features of the sensitivity study.

\subsection{Playground data}
We use data from the third \ac{LVK} observing run (O3) as playground data to assess the sensitivity of the cWB-GMM-JSD search. The O3 run consisted of two phases separated by a one-month hiatus - O3a from 1st April 2019 to 1st October 2019, and O3b from 1st November 2019 to 27th March 2020. Henceforth, we refer to the data from the 3-detector network consisting of \ac{LLO}, \ac{LHO} and Virgo as LHV data and the data from the 2-detector network with \ac{LLO} and \ac{LHO} detectors as LH data.  The strain data for both signals and noise are sampled at 4096 Hz. \ac{cWB} processes the strain data based on the detector combination and produces triggers as described in Sec. \ref{cwb-gmm-overview}, with each trigger characterized by a set of attributes. We further process the \ac{cWB} attributes for these two networks separately using the \ac{GMM} approach as outlined in Sec. \ref{cwb-gmm-overview}. For \ac{PE}, we use the strain data from the detectors, with the \ac{PSD} calculated from data around the time of the trigger. 

\subsubsection{Signal Injections}

We consider a set of \textit{ad-hoc} waveforms which are commonly used to assess the sensitivity of weakly-modeled all-sky short pipelines: \acf{WNBs}, \acf{SGs} and \acf{GAs}. These waveforms embody the basic features of the target signals of the all-sky short search - short duration, pulse-like morphology and limited bandwidth. We also consider astrophysically-motivated simulations of \ac{CCSN} for the sensitivity study. \ac{CBC} waveforms \cite{imrphenomxas-ref} are used for carrying out the parameter estimation needed to compute $\mathcal{J}$, using the same priors as in \cite{Ghosh_2023}. The details of the injected signals considered for the sensitivity study are seen below:
\begin{enumerate}
    \item \acf{GAs} are parameterised by the amplitude $\mathcal{A}$ and the one-sigma time $\tau$. We choose $\mathcal{A}$ in such a way that the root sum squared amplitude $h_{rss}$ (defined in Eqn. \ref{eq:hrss}) takes values of the form $(\sqrt{3})^N \times 5 \times 10^{-23} \mathrm{Hz}^{-\frac{1}{2}}$ with $N$ ranging from 0 to 8. We list the $\tau$ values in Table \ref{tab:ad-hoc_def}.
    \item \acf{SGs} are parameterised by the amplitude $\mathcal{A}$, the quality factor $Q$ and the duration $\tau$. We follow the same distribution for $\mathcal{A}$ as \ac{GAs}. We list the $Q$ and $\tau$ values in Table \ref{tab:ad-hoc_def} 
    \item \acf{WNBs} are parameterised by $h_{rss}$ (chosen from the same distribution as the previous two), a  lower frequency $f_{low}$, a bandwidth $\Delta f$, and duration $\tau$. We list these parameters also in Table \ref{tab:ad-hoc_def}.
    \item Core collapse supernovae (CCSN) signals: We inject the same sets of simulations that were reported in \cite{Smith_2024}. We take 10 neutrino-driven explosion models - Andresen et al. 2017 \cite{Andresen:2017s11} (And s11), M$\ddot{u}$ller et al. 2012 \cite{Muller:2012L15} (Mul L15), Kuroda et al. 2016 \cite{Kuroda:2016SFHx} (Kur SFHx), O’Connor $\&$ Couch 2018 \cite{OConnor:2018-m20} (Oco mesa20), Powell $\&$ M$\ddot{u}$ller 2019 \cite{Powell:2018-s18} (Pow he3.5, s18), Radice et al. 2019 \cite{Radice:2018-s9} (Rad s9, s13, s25), and 1 magnetorotationally-driven explosion model: Abdikamalov et al. 2014 \cite{Abdikamalov:2013sta}. 
\end{enumerate}
All simulated waveforms that are used for the sensitivity study are injected into artificial noise generated from the detectors' \ac{PSD} for the O3 observing run.  

\begin{table}[h]
    \centering
    \begin{tabular}{c c c}
         \hline 
         \hline
          & Gaussian Pulse (GA) & \\
         \hline

          &  & $\tau$ (ms) \\
         \hline
 
           & & 0.1   \\
           & & 1   \\
           & & 2.5   \\
           & & 4  \\
         \hline 
          & Sine-Gaussian (SG) & \\
         \hline

          $f_0$ (Hz) & $Q$ &  \\
          \hline
 
          70 &  3 &  \\
          70 &  9 &  \\ 
          70 & 100 &  \\
          100 & 9 &  \\
          153 & 9 &  \\
          235 & 3 &  \\
          235 & 9 &  \\ 
          235 & 100 &  \\
          361 & 9 &  \\
          554 & 9 &  \\
          849 & 3 &  \\
          849 & 9 &  \\
          849 & 100 &  \\ 

         \hline 
          & White Noise Burst (WNB) &  \\ 
         \hline

          $f_{low}$ (Hz) & $\Delta f$ (Hz) & $\tau$ (s) \\ 
          \hline
 
          150 & 100 & 0.1 \\
          300 & 100 & 0.1 \\
          750 & 100 & 0.1 \\
          \hline
          
    \hline
    \hline
         
    \end{tabular}
    \caption{Table of generic ad-hoc simulations with defining parameters used in the O3 all-sky short search.}
    \label{tab:ad-hoc_def}
\end{table}

\subsubsection{Background data}
To assess the statistical significance of a \ac{GW} signal, it is crucial to obtain an estimate of the noise distribution, that is, to obtain the distribution of chance-coincident noise triggers for a given search, known as the \textit{background}. cWB uses the \textit{time-slide} method to generate such a background.
In the time-slide approach, the data in one detector is slid along the time axis with respect to other detectors by time-shifts which are outside the light travel times between the detectors. This process ensures that when the search algorithm runs on the background data, all triggers are of noise origin. In this work we do not perform time-sliding ourselves, but rather post-process the background produced by the O3 \ac{cWB} search \cite{xgboost_2023}. For the remainder of this work, we use \textit{background triggers} and \textit{noise triggers} interchangeably. 

\begin{figure}[htb!]
    \begin{subfigure}{.48\textwidth}
     \centering
     \includegraphics[width=\linewidth]{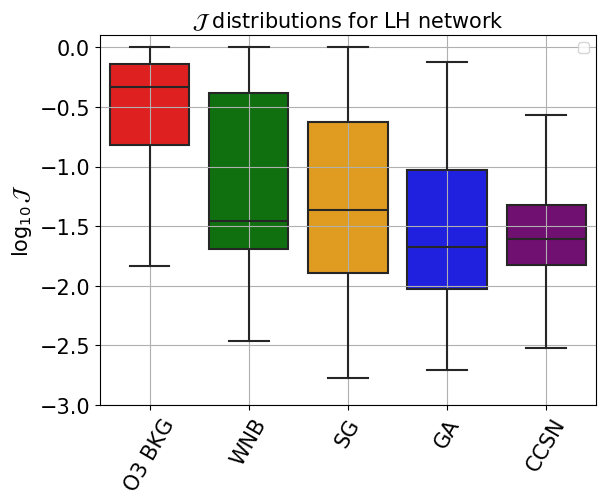}
    \end{subfigure}

    \begin{subfigure}{.48\textwidth}
     \centering
     \includegraphics[width=\linewidth]{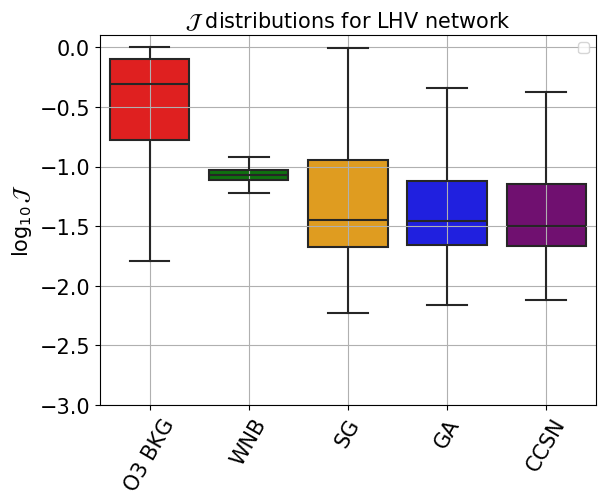}
    \end{subfigure}
    \caption{The multi-detector network JSD ($\mathcal{J}$) distributions for background and signal triggers. The top panel shows the distributions for the LH network, and the bottom panel shows the distributions for the LHV network. \footnote{These are standard box-whisker plots, in which the boxes show the first and third quartiles along with the median, and the ends of the ``whiskers'' show the extreme ends of the distribution.}}
    \label{J-dist-LH-LHV}
\end{figure}
 
\subsection{Reweighing the detection statistic}

\subsubsection{Reweighing the signals}
\ac{cWB} simulated $\sim 100,000$ signals across all waveform types for each of O3a and O3b, which were used for testing the sensitivity of cWB-based pipelines. However, for computing the multi-detector network JSD, we take a sample distribution of 100 injections of each waveform type listed in Table \ref{tab:ad-hoc_def} from the same distribution from which \ac{cWB} draws its signals, and calculate their $\mathcal{J}$ distribution. 

Fig. \ref{J-dist-LH-LHV} shows the distributions of $\mathcal{J}$ for different ad-hoc injections, \ac{CCSN} simulations and background triggers, for both the LH and LHV networks. 
For most signals, $\mathcal{J}$ is small compared to unity. For each of the signal types listed in \ref{tab:ad-hoc_def}, and for each \ac{CCSN} simulation, we use the median value of the $\mathcal{J}$ distribution to compute $T_{\mathcal J}$. There are certain signals for which $\mathcal{J}$ is too high and this approximation breaks, but they comprise only a small fraction of the entire signal distribution.

\subsubsection{Reweighing the background triggers}
The background triggers are noisy transients occurring in different detectors, but temporally coincident due to time-sliding. For most searches, modeled or otherwise, the loud noise triggers of the background distribution limit the search sensitivity. In this work, we down-rank each background trigger above the threshold $T_0 = 0$ and thus mitigate the effect of loud glitches on the sensitivity of the search as discussed in the next section. In our study, the loudest glitches are mostly blip-like, with short-duration, limited bandwidth and appear as pulses in time-frequency representations \cite{Cabero:2019blip}.


We also observe in Fig. \ref{J-dist-LH-LHV} that background triggers typically have much larger $\mathcal{J}$ values than the ad-hoc and \ac{CCSN} simulations. Thus, $\mathcal{J}$ is not merely a discriminator between \ac{IMBH} binary signals and noise triggers as was shown in \cite{Ghosh_2023}, but also between generic ad-hoc signals and noise triggers. This broadens its implementability from targeted massive \ac{BBH} searches to all searches for short-duration signals, and is particularly suitable for burst searches, such as \ac{cWB}-based searches. 

\subsection{Statistical significance estimation}
Using the background, we estimate the statistical significance of an injection by computing the associated \acf{FAR}, which is the number of false alarms (noise triggers louder than the given trigger) per year. For any astrophysical injection with value of ranking statistic $T_J = T_J^*$, the \ac{FAR} is:
\begin{equation}\label{eqn:far-defn}
    \mathrm{FAR} = \frac{n_{\mathrm{BKG}}(T_J \geq T_J^*)}{t_{\mathrm{BKG}}} \ ,
\end{equation}
where $n_{\mathrm{BKG}}(T_J \geq T_J^*)$ is the number of background triggers with ranking statistic $T_J \geq T_J^*$, and $t_{\mathrm{BKG}}$ is the total background time available for the analysis. An equivalent measure of significance is the \acf{IFAR}, which is the reciprocal of \ac{FAR}, and is the duration in years in which we expect to see a noise trigger as loud as the given trigger.

We apply the same procedure to assess the statistical significance for the simulations as well as the search for \ac{GW} candidates in the observational data. During the search, candidates with \ac{FAR} below a pre-determined threshold (\ac{IFAR} above a corresponding pre-determined threshold) are flagged as potential \ac{GW} events.  

\subsection{Sensitivity measures}

After assigning a \ac{FAR} value to each injection using Eqn. \ref{eqn:far-defn}, we estimate the sensitivity of the search to different types of signals. For different \ac{IFAR} thresholds, we compare the sensitivity of the new search (cWB-GMM-JSD) with the previous version from \cite{Smith_2024} (cWB-GMM). We use two sensitivity measures - amplitude at which $50\%$ of the simulated signals are recovered, denoted by $h_{rss50}$, and sensitive volume $\mathcal{V}$ \cite{allsky-o1,allsky-o2,allskyo3}. 

The  $h_{rss}$ is the root sum square of the GW strain: 

\begin{equation}
    h_{rss} = \sqrt{\int_{-\infty}^{\infty} \Big(h_{+}^2(t) + h_{\times}^2(t) \Big)dt}
    \label{eq:hrss}
\end{equation} 

where $h_{+}$ and $h_{\times}$ are the plus and cross polarisations of the GW signal. We calculate the detection efficiency of a given waveform as a function of $h_{rss}$, which is the fraction of detected events at a given false alarm threshold over the number of injected events for injected $h_{rss}$ amplitude values. The $h_{rss50}$ statistic is the $h_{rss}$ amplitude at which 50$\%$ detection efficiency is achieved. Smaller $h_{rss50}$ indicates a greater ability to detect weaker signals. 

Assuming standard siren sources distributed uniformly in space, the sensitive volume is obtained by integrating the detection efficiency as a function of distance over all of space \cite{xgboost_2023}:
\begin{equation}
    \mathcal{V} = 4\pi (r_0 h(r_0))^3 \int_0^{\infty} \frac{dh}{h^4} \epsilon (h) \,
\end{equation}
where $h(r_0)$ is the $h_{rss}$ value at some reference distance $r_0$, and $\epsilon(h)$ is the detection efficiency as a function of $h_{rss}$ ($rss$ has been dropped for brevity). At the same $h_{rss}$, cWB-GMM and cWB-GMM-JSD have different detection efficiencies - $\epsilon_{\mathrm{cWB-GMM}}(h)$ and $\epsilon_{\mathrm{cWB-GMM-JSD}}(h)$. Hence, the ratio of their sensitivity volumes becomes: 
\begin{equation}
    \frac{\mathcal{V}_{\mathrm{cWB-GMM-JSD}}}{\mathcal{V}_{\mathrm{cWB-GMM}}} = \frac{\int_0^{\infty} \frac{dh}{h^4} \epsilon_{\mathrm{cWB-GMM-JSD}} (h)}{\int_0^{\infty} \frac{dh}{h^4} \epsilon_{\mathrm{cWB-GMM}} (h)}
\end{equation}

\section{Results} \label{Sec:results}

In this section, we show the $\mathcal{J}$ distributions for signals and background triggers and the sensitivity comparison between the new cWB-GMM-JSD search and the old cWB-GMM search. We further run this search on the O3 foreground data and present the results of our search.

\subsection{$\mathcal{J}$ distributions for ad-hoc waveforms and \ac{CCSN} simulations}

Figure \ref{J-dist-LH-LHV} shows the $\mathcal{J}$ distributions for all signal types (GA, SG, WNB and CCSN) and background triggers for both the LH and the LHV network. We observe that all four types of signals have much lower $\mathcal{J}$ values than background triggers. This proves our assertion that even if a \ac{CBC} model is used for \ac{PE} of \ac{GAs}, \ac{SGs}, \ac{WNBs} and \ac{CCSN}, the single-detector posteriors are largely consistent with one another. Although the bulk of the distribution for each signal type lies below $0.1$, there are still a few cases in which the $\mathcal{J}$ values are high. 

\begin{figure*}
    \centering
    \includegraphics[width=\linewidth]{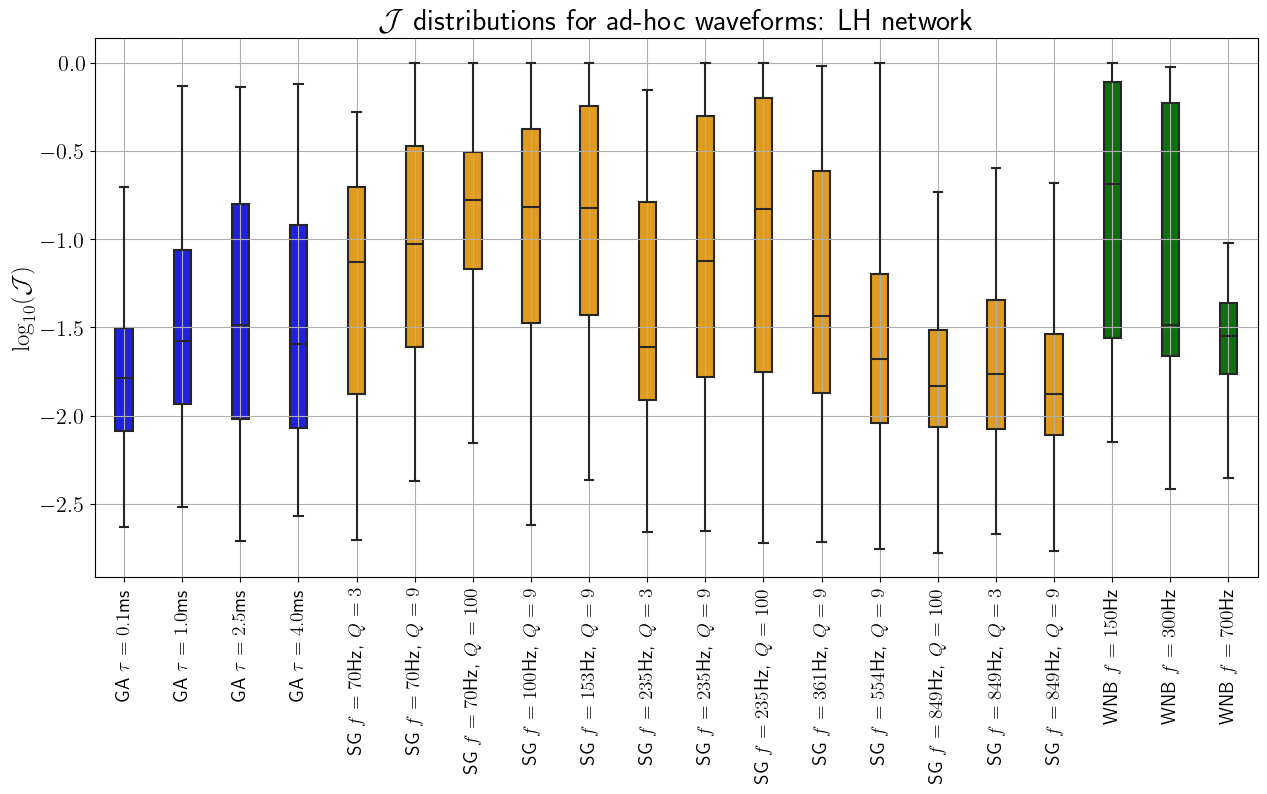}
    \includegraphics[width=\linewidth]{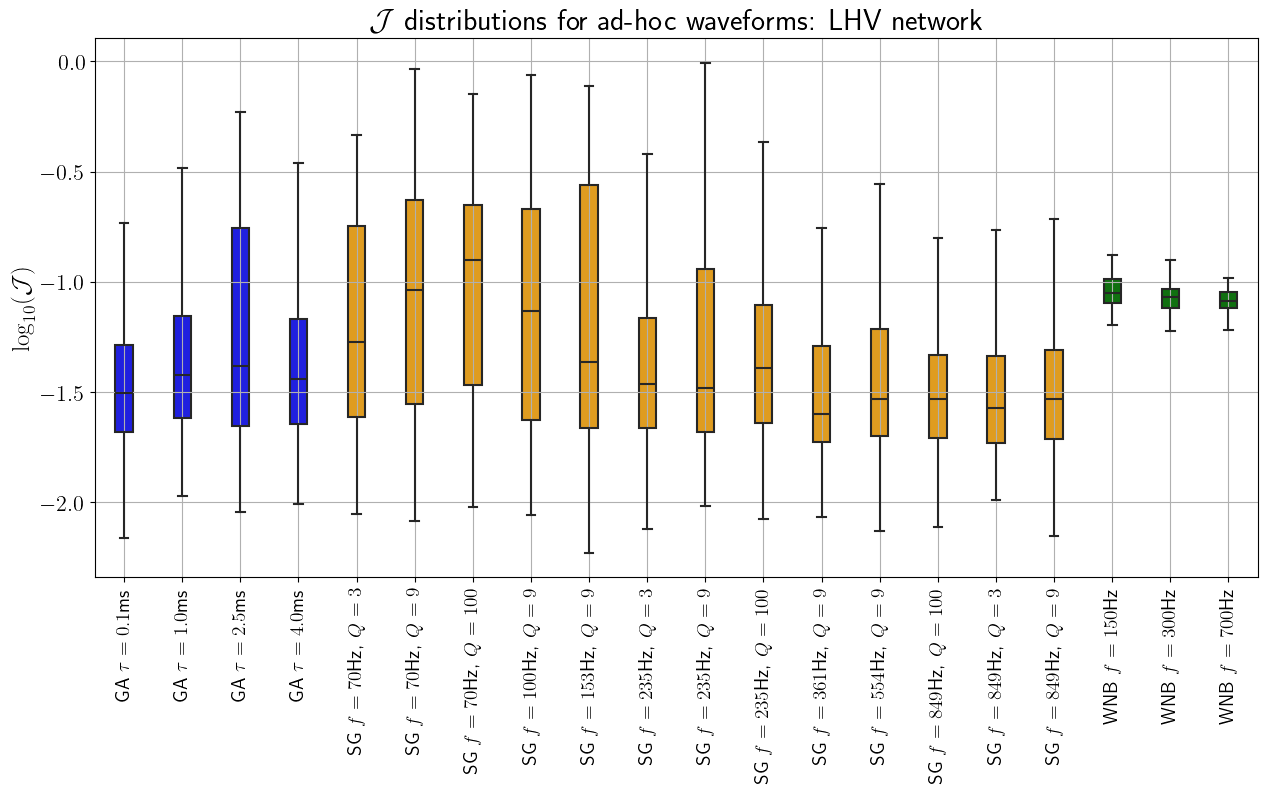}
    \caption{Distributions of $\mathcal{J}$ for different ad-hoc waveforms for the LH network (top panel) and the LHV network (bottom panel). The y-axis shows $\mathcal{J}$ in the log scale. The x-axis shows the waveforms. The different waveform groups are represented by different colors. The \ac{GAs} are shown in blue, the \ac{SGs} are shown in orange, and the \ac{WNBs} are shown in green.}
    
    \label{J_vs_adhoc}
\end{figure*}

Figure \ref{J_vs_adhoc} shows the variation of the $\mathcal{J}$ distributions for \ac{SGs}, \ac{GAs} and \ac{WNBs} with injected parameters. In this figure, the y-axis shows $\mathcal{J}$ in the log-scale and the x-axis labels the different ad-hoc waveforms.  We make the following observations for the different waveforms:
\begin{enumerate}
    \item \ac{GAs}: More than $75 \%$ of the $\mathcal{J}$ distribution lies below 0.1 for $\tau = 0.1 \mathrm{ms}$ and $\tau = 1.0 \mathrm{ms}$. Higher $\tau$ values have longer tails. In particular, for $\tau = 2.5 \mathrm{ms}$, more than $25 \%$ of the signals are above $\mathcal{J} = 0.1$.  
    \item \ac{SGs}: we see that the lower-frequency signals (70Hz, 100Hz, 153Hz and 235Hz) generally tend to have higher values of $\mathcal{J}$ than the higher-frequency signals (361Hz, 554Hz, 849Hz). Among the lower-frequency signals, the ones with higher $Q$ values ($Q = 100,9$) have higher $\mathcal{J}$ than lower $Q$ waveforms. In short, signals with low frequency and high $Q$ are more prone to have high $\mathcal{J}$ than other Sine-Gaussian waveforms.  
    \item \ac{WNBs}: Waveforms with higher $f_{low}$ tend to have lower $\mathcal{J}$.
\end{enumerate}

\begin{figure}[h]
    \centering
    \includegraphics[width=0.95\linewidth]{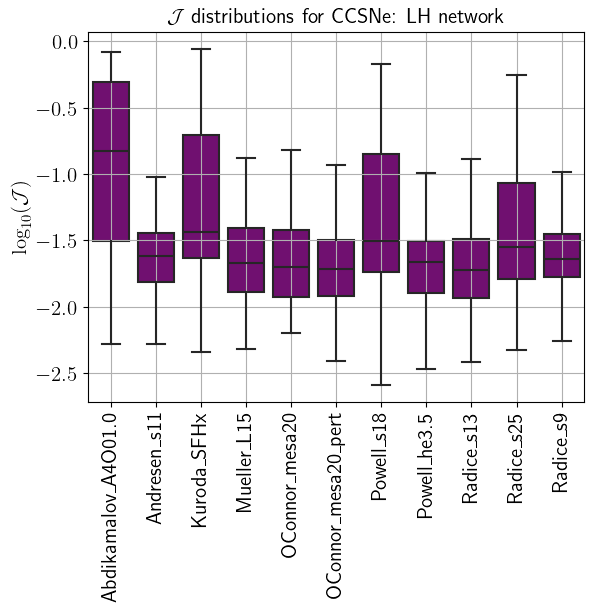}
    \includegraphics[width=0.95\linewidth]{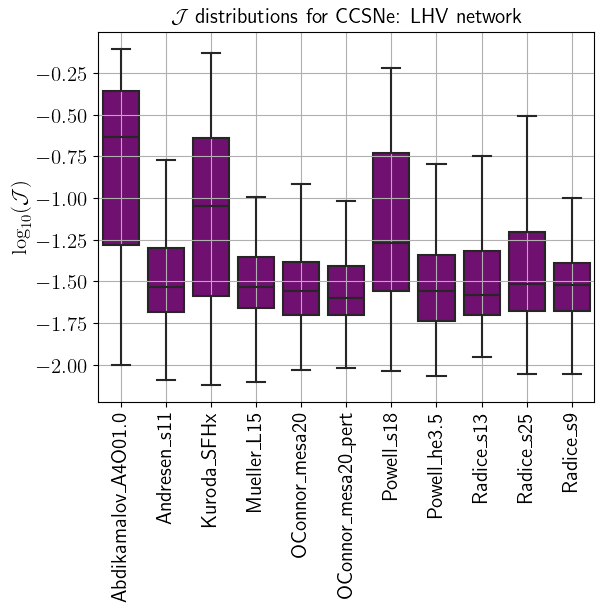}
    \caption{Distributions of $\mathcal{J}$ for different \ac{CCSN} simulations for the LH network (top panel) and the LHV network (bottom panel). The y-axis shows $\mathcal{J}$ in the log scale. The x-axis shows the simulations.}
    

    \label{fig:J_vs_CCSN}
\end{figure}
Figure \ref{fig:J_vs_CCSN} shows $\mathcal{J}$ values for each CCSN waveform. The x-axis labels the different simulations. For both the LH and the LHV networks, we make the following observations.
\begin{itemize}
    \item For the \texttt{Abdikamalov\_A4O01.0} simulations, more than $50 \%$ of the signals lie above $\mathcal{J} = 0.1$. This causes signals of this type to be down-ranked materially, consequently leading to a degradation in the sensitivity to this waveform type as discussed in the next subsection. 
    \item For the \texttt{Kuroda\_SFHx}, and \texttt{Powell\_s18} simulations, more than $25 \%$ of the waveforms have $\mathcal{J} \geq 0.1$.  
    \item For the other \ac{CCSN} simulations, $\mathcal{J}$ is typically less than $0.1$
\end{itemize}

The lower $\mathcal{J}$ values of the high-frequency signals and the higher $\mathcal{J}$ values of the low-frequency signals can be explained in the following way. For the \ac{CBC} waveforms which we are using for \ac{PE}, lower-mass waveforms are longer and have higher frequencies, whereas higher-mass waveforms are shorter and have lower frequencies. So, when we fit \ac{CBC} waveforms to short-duration, high-frequency waveforms, the parameter estimates have a high uncertainty because no \ac{CBC} waveform fits the signal perfectly. As a result of this high uncertainty, the posteriors are broad in all detectors and cause the $\mathcal{J}$ values to be lower. For low-frequency short-duration signals, the posteriors have sharper peaks, but the means of the posteriors do not always coincide exactly between detectors, causing the $\mathcal{J}$ values to be slightly higher than the high-frequency signals. 

The extremely low $\mathcal{J}$ values for most of the \ac{CCSN} simulations can be explained by the fact that the morphologies of these waveforms are distinctly different from those of the short-duration chirping \ac{CBC} waveforms which we are using for the \ac{PE}, and hence the posteriors are very broad and mostly uninformative. 

In summary, a minority of ad-hoc and CCSN waveforms have $\mathcal{J}$ values $\geq 0.1$. For these signals, the approximation $T_J \approx T$ will break, but for other waveforms the approximation is perfectly valid. The sensitivity improvement for the majority of waveforms outweighs the potential loss in sensitivity for the few waveforms where the approximation is not valid.

\subsection{Improving cWB-GMM with $\mathcal{J}$}
\label{cwb-gmm-jsd-in-action}

\begin{figure*}
    \centering
    \includegraphics[width=\linewidth]{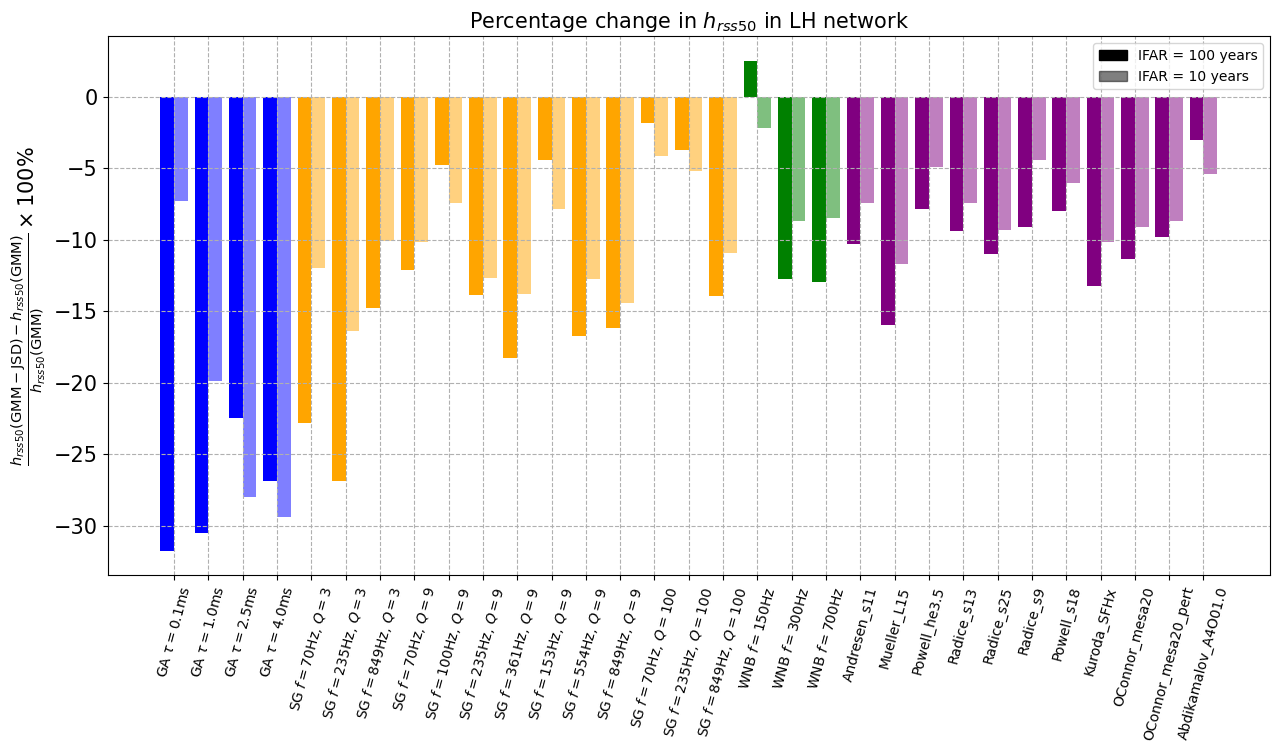}
    \includegraphics[width=\linewidth]{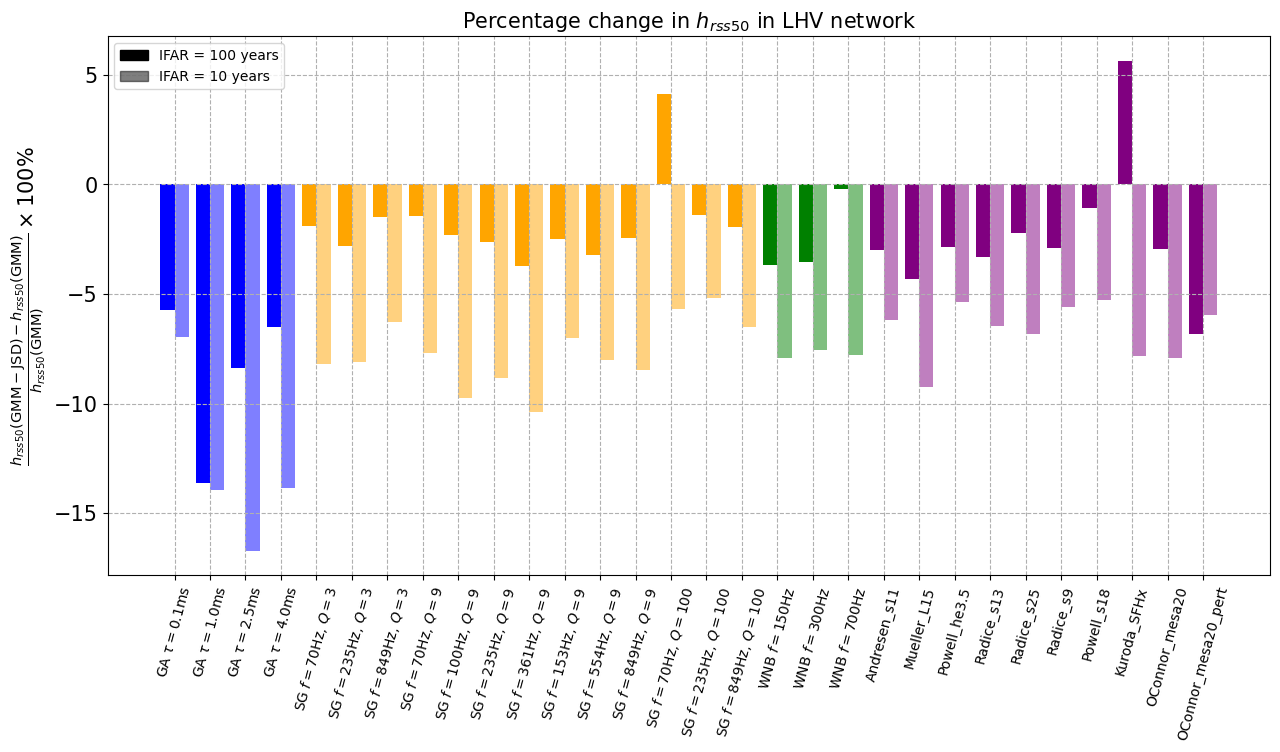}
    \caption{Percentage differences between $h_{rss50}$ estimates for cWB-GMM and cWB-GMM-JSD with the LH (\ac{LLO}-\ac{LHO}) network (top) and the LHV (\ac{LLO}-\ac{LHO}-Virgo) network across all ad-hoc and CCSN waveforms. A decrease in $h_{rss50}$ indicates increased sensitivity. Estimates at IFAR$\geq$10 years are shown in lighter shades, while the estimates at IFAR$\geq$100 years are displayed in darker shades. For the LHV network, the Abdikamolov CCSN waveform (\textit{Abd A4O01.0}) does not achieve 50\% detection efficiency within the injected $h_{rss}$ range, thus is set to the maximum injected $h_{rss}$ of 4$\times10^{-21}$Hz$^{-1/2}$, leading to a huge percentage increase in the $h_{rss50}$ that is off the scale, and hence not shown in the plot.}
    \label{hrss50_o3ab}
\end{figure*}

We now showcase the results of integrating the $\mathcal{J}$ statistic with the cWB-GMM pipeline. Fig. \ref{hrss50_o3ab} shows comparisons between the $h_{rss50}$ values for cWB-GMM and cWB-GMM-JSD for the LH and LHV networks in terms of percentage changes. A decrease in the $h_{rss50}$ value indicates the ability of the search to detect signals at smaller amplitudes, and hence signifies an increase in sensitivity. An \ac{IFAR} of 100 years is generally taken to be the threshold for significant detection for all-sky searches \cite{allsky-o1,allsky-o2,allskyo3}. Here, we compute the $h_{rss50}$ values at \acp{IFAR} of 10 years and 100 years.

\subsubsection{Sensitivity improvements for the LH network}
For the LH network, the improvements in $h_{rss50}$ due to the \ac{JSD} post-processing are as follows:
\begin{itemize}
    \item \ac{GAs}: At \ac{IFAR} = 100 (10)years, $h_{rss50}$ decreases by $\sim 20-30\%$ ($\sim 7 - 30 \%$). This translates roughly to an increase in the sensitivity volume by a factor of $\sim 1.5 - 3$ ($\sim 1.3 - 1.5$) In terms of percentage reduction in $h_{rss50}$, \ac{GAs} show the best improvement among all the waveform types. This can be attributed to the fact that the $T$ and $T_J$ distributions of the \ac{GAs} have a higher support near the loudest background triggers as compared to the other waveform types. Thus, the re-weighting causes more \ac{GAs} to be boosted in significance as compared to the other waveforms. For a detailed explanation, see App. \ref{explaining-sensitivity-improvements}
    \item \ac{SGs}: At \ac{IFAR} = 100 (10) years, $h_{rss50}$ decreases by $\sim 5 - 20 \%$ ($\sim 5 - 15 \%$). These changes in $h_{rss50}$ translate roughly to an increase in the sensitivity volume  by a factor of $\sim 1.05 - 1.5$ at both IFARs.    
    \item \ac{WNBs}: At \ac{IFAR} of 100(10) years, $h_{rss50}$ decreases by $\sim 12\%$ ($\sim 8\%$) for the waveforms with $f_{low} = 300$Hz and $f_{low} = 700$Hz. These changes translate roughly to an increase in the sensitivity volume by a factor of $\sim 1.2 - 1.3$ at both IFARs. For the waveforms with $f_{low} = 150$Hz, we see a slight increase ($\sim 2\%$) in $h_{rss50}$ at \ac{IFAR} = 100 years, and only a slight decrease ($\sim 2\%$) at \ac{IFAR} = 10 years. These correspond to the new sensitivity volume being 0.95 and 1.05 times that of the the old one at IFAR = 100 years and 10 years. This can be attributed to the higher values of $\mathcal{J}$ for this waveform type, as can be seen in Fig. \ref{J_vs_adhoc}. The higher values of $\mathcal{J}$ are counteracting the effect of reduction in background, leading to a negligible change in the sensitivity. 
    \item \ac{CCSN}: At \ac{IFAR} = 100 (10) years $h_{rss50}$ decreases by $\sim 7 - 15 \%$ ($\sim 5 - 10 \%$) for all the simulations except the \texttt{Abdikamalov\_A4O01.0} waveform, for which the percentage improvement is around $5 \%$ or less at both \ac{IFAR}s. For most of the CCSN simulations, the percentage decreases in $h_{rss50}$ correspond to an increase in the sensitivity volume by a factor of $\sim 1.1 - 1.3$ at both IFARs. The exceptional behaviour of \texttt{Abdikamalov\_A4O01.0} again can be ascribed to the higher values of $\mathcal{J}$ for this simulation type, as can be seen in Fig. \ref{fig:J_vs_CCSN}
\end{itemize}

\subsubsection{Sensitivity improvements for the LHV network}
For the LHV network, the improvements in $h_{rss50}$ for the ad-hoc and \ac{CCSN} waveforms are as follows:
\begin{itemize}
    \item \ac{GAs}: At both \ac{IFAR} = 100 years and \ac{IFAR} = 10 years $h_{rss50}$ decreases by $\sim 5-15 \%$. For the LHV network also, the performance of \ac{GAs} is better than the other waveforms, for the same reasons as the LH case, and are described in Appendix \ref{explaining-sensitivity-improvements}.
    \item \ac{SGs}: At \ac{IFAR} = 100 (10) years, the percentage reduction in $h_{rss50}$ is $< 5 \%$ ($5 - 10 \%$) for all the waveforms, except $Q = 100, f = 70 \mathrm{Hz}$, for which there is an increase in $h_{rss50}$ by $4\%$ at \ac{IFAR} = 100 years. 
    \item \ac{WNBs}: At \ac{IFAR} = 100 (10) years, the percentage decrease in $h_{rss50}$ is $<5\%$ ($\sim 7\%$).  
    \item \ac{CCSN}: For most simulations, at \ac{IFAR} = 100 (10) years, the percentage improvements in $h_{rss50}$ are $< 5 \%$ ($5-7 \%$). The only exceptions to this are \texttt{Abdikamalov\_A4O01.0} and \texttt{Kuroda\_SFHx}. For \texttt{Abdikamalov\_A4O01.0}, at \ac{IFAR} = 100 years, $50\%$ detection efficiency is not reached, which prompts the algorithm to set the maximum injected  $h_{rss}$ as the $h_{rss50}$ value for this waveform (this waveform is not shown in Fig. \ref{hrss50_o3ab}). For \texttt{Kuroda\_SFHx}, $h_{rss50}$ increases by $\sim 6\%$ at \ac{IFAR} = 100 years.
\end{itemize}

For the LHV network, the increase in the sensitivity volume is less than $\sim 1.1$ for all the waveform types.


\subsection{Search for \ac{GW} events in O3 data with cWB-GMM-JSD}
\label{re-analysis}

In this subsection, we discuss the results of the search for \ac{GW} events in O3 data with the cWB-GMM-JSD algorithm. We compare the significance of the detected events with the results of the cWB-GMM search reported in \cite{Smith_2024}. We tabulate this comparison in Table \ref{combined_table}.

No new event was observed with cWB-GMM-JSD. For the LH network, the same candidates that were reported in \cite{Smith_2024} are observed, albeit with revised statistical significances. For the majority of events, there is a notable increase in \ac{IFAR}, indicating that events are detected with higher significance. For instance, the GW190521 event, a confident \ac{IMBH} binary event \cite{gw190521-discovery,gw190521-properties} was initially detected by cWB-GMM with an IFAR of less than 4 years and  is now assigned an IFAR of approximately 16 years by cWB-GMM-JSD. The events with reduced statistical significance are GW190412 and GW191109\_010717. For GW190412, the high $\mathcal{J}$ value comes mainly from the differences in the $\chi_{\mathrm{eff}}$ posteriors in the \ac{LLO} and \ac{LHO} detectors. The new IFAR of 20 years is less compared to its earlier value of 65 years, but it is still an acceptable value for confident detection. For GW191109\_010717, it is well-known that there were glitches in both LIGO detectors caused by the scattering of light \cite{gw191109-properties}, and the new method down-ranks the event because of the presence of glitches.  

Four GW candidates are detected (with \ac{IFAR} $\geq$ 1 year) with the LHV network, similarly to the previous results: GW200224\_222234, GW190412, GW190828\_063405 and GW190706\_222641. The \ac{IFAR} estimates are significantly increased for two of these events. 


For events which are observed with improved significance in both the LH and the LHV networks, the \ac{IFAR} increases by a factor of $\sim 1.3 - 5$. 
\begin{table*}
    \centering
    \setlength{\tabcolsep}{5.5pt}
    \begin{tabular}{c | c c c c | c c c c}
         \hline
         \hline

          \multirow{3}{*}{Event Name } & \multicolumn{4}{c|}{LH network} & \multicolumn{4}{c}{LHV network}  \\ 
            & \multirow{2}{*}{$T$} & cWB-GMM  & \multirow{2}{*}{$T_J$}  & cWB-GMM-JSD & \multirow{2}{*}{$T$} & cWB-GMM  &  \multirow{2}{*}{$T_J$} & cWB-GMM-JSD \\
            &  & IFAR ({\it yr}) &  & IFAR ({\it yr}) &  & IFAR ({\it yr}) &  & IFAR ({\it yr}) \\
          \hline

          GW200224\_222234 & 21.86 & 109.62 & 18.72 & 219.24 & 18.70 & 10.80 & 15.39 & 29.7 \\
          GW190521\_074359 & 32.37 & 98.09 & 22.47 & 196.19 & -5.54 & 0.06 & -5.54 & 0.06 \\ 
          GW190412 & 23.38 & 65.40 & 6.37  & 19.62 & 17.41 & 19.10 & 4.85 & 9.55 \\
          GW190519\_153544 \footnote{For LHV, this event was obtained from the ``extended segments" in cWB, for which the minimum analysis segment time is reduced.} & 19.61 & 39.24 & 17.77  & 196.19 & -2.04 & 0.45 & -2.04 & 0.45 \\
          GW191204\_171526 & 13.05 & 31.32 & 13.05  & 109.62 & - & - & - & - \\
          GW191109\_010717 & 9.63 & 12.18 & 1.95 & 2.71 & - & - & - & - \\
          GW190828\_063405 & 10.68 & 11.54 & 8.57  & 39.24 & 5.10 & 6.88 & 2.72 & 5.55 \\
          GW190706\_222641 & 7.55 & 4.67 & 6.05   & 17.84 & 7.35 & 1.70 & 3.87 & 8.19  \\
          GW200311\_115853 & 6.35 & 4.22 & 3.61 & 9.53 & -0.24 & 0.03 & -0.24 & 0.03 \\
          GW190521 & 6.49 & 3.77 & 5.35  & 16.35 & -5.54 & 0.06 & -5.54 & 0.06  \\
          GW190408\_181802 & 5.36 & 2.48 & 3.05  & 6.77 & -3.71 & 0.18 & -3.71 & 0.18 \\
          GW191222\_033537 & 3.84 & 1.49 & 1.99 & 2.85 & - & - & - & -\\
          GW200225\_060421 \footnote{For LH, this event was obtained from the ``extended segments" in cWB, for which the minimum analysis segment time is reduced.} & 3.09 & 1.06 & 1.02 & 1.09 & - & - & - & - \\
          GW190915\_235702 & 2.96 & 1.05 & 1.27  & 1.98 & -10.55 & 0.004 & -10.55 & 0.004 \\

         \hline
         \hline
    \end{tabular}
      \caption{Table detailing the results of cWB-GMM and cWB-GMM-JSD on O3 data. For each detector network (LH and LHV), the following information are shown - $T$ value, \ac{IFAR} value obtained from cWB-GMM, $T_J$ value and \ac{IFAR} value obtained from cWB-GMM-JSD. GW events detected with IFAR$\geq$1 year in the LH network are displayed.}
      \label{combined_table}
\end{table*}

\section{Conclusions}
\label{Sec:conclusions}

The presence of non-stationary, non-Gaussian noise transients or glitches in the strain data of ground-based \ac{GW} detectors has always been a hindrance to the detection of astrophysical signals. The morphological similarities between signals and certain types of glitch classes, such as Blips, aggravates the problem. The search for new signal-noise discriminators and their integration with existing search algorithms is an area of active research. In this work, we developed a \ac{JSD}-based signal-noise discriminator for a multi-detector network which assesses the dissimilarity between the posterior distributions averaged over all  astrophysical parameters for all possible detector pairs. This network based mathematical construct referred to as multi-detector network JSD $\mathcal{J}$, takes  into account the differences in sensitivities of the detectors in terms of the noise power spectral density and can be applied to an arbitrary number of detectors with different sensitivities. We demonstrate that the multi-detector network JSD can be used to distinguish between loud noise triggers and generic ad-hoc signals like \acf{GAs}, \acf{SGs} and \acf{WNBs}, in addition to various astrophysically motivated simulations of \ac{CCSN} signals. We seamlessly integrate this multi-detector network JSD in revising the ranking statistic in such a way that it remains almost unchanged for most of the signals, but heavily penalizes the background triggers - thereby leading to an increase in the sensitivity. This has been demonstrated with data from the third observing run of LIGO and Virgo (O3) and for the cWB-GMM pipeline - one of the leading search algorithms used by the collaboration for all-sky searches. With the \ac{JSD}-based post-processing, the upgraded search algorithm, termed cWB-GMM-JSD shows better sensitivities than cWB-GMM to most ad-hoc waveforms and \ac{CCSN} signals at IFARs of 10 years and 100 years. Among the ad-hoc waveforms, the \ac{GAs} show the best improvement in terms of percentage reduction of $h_{rss50}$, which is around $30 \%$ for the LH network and $5-15 \%$ for the LHV network. For the other waveforms, the percentage reduction in $h_{rss50}$ is around $10-20\%$ ($5-10 \%$) for the LH network (LHV network). For the \ac{WNBs} and the \texttt{Abdikamalov\_A4O01.0} \ac{CCSN} simulation, there is a deterioration in the sensitivity at \ac{IFAR} = 100 years for the LHV network. The reweighing approach derived from the parameter consistency across the multi-detector network proposed in this work is first of its kind, generic in nature, and independent of the search algorithm, making it implementable in most \ac{GW} searches.

\begin{acknowledgments}

The authors would like to thank the authors of \cite{xgboost_2023} for the production of cWB triggers used for the analysis in this paper. The authors would like to thank Shubhanshu Tiwari, Edoardo Milotti and Marco Drago for interesting discussions and suggestions.
The authors acknowledge the computational resources which aided the completion of this project, provided by LIGO-Laboratory and support by the National Science Foundation (NSF) Grants No.PHY-0757058 and No.PHY-0823459. This research has made use of data or software obtained from the Gravitational Wave Open Science Center (gwosc.org), a service of the LIGO Scientific Collaboration, the Virgo Collaboration, and KAGRA. This material is based upon work supported by NSF's LIGO Laboratory which is a major facility fully funded by the National Science Foundation.
SG acknowledges fellowship support from MHRD, Government of India. 
L.S acknowledges support from the European Union - Next Generation EU Mission 4 Component 1 CUP J53D23001550006 with the PRIN Project No. 202275HT58, and by ICSC – Centro Nazionale di Ricerca in High Performance Computing, Big Data and Quantum Computing, funded by European Union – NextGenerationEU.
JS is supported in part by the National Research Foundation of Korea (NRF) funded by the Ministry of Education (NRF-2022R1I1A207366012).
AP acknowledges the support from SPARC MoE grant SPARC/2019-2020/P2926/SL, Government of India.
ISH was supported by Science and Technology Facilities Council (STFC) grants ST/V001736/1 and ST/V005634/1. 
GV acknowledge the support of the National Science Foundation under grant PHY-2207728?
\\
\end{acknowledgments}

\appendix

\section{Understanding differences between the percentage improvements in sensitivity across waveform types.}{\label{explaining-sensitivity-improvements}}

\begin{figure*}{\label{T-TJ-dist}}
    \centering
    \includegraphics[width=0.45\textwidth]{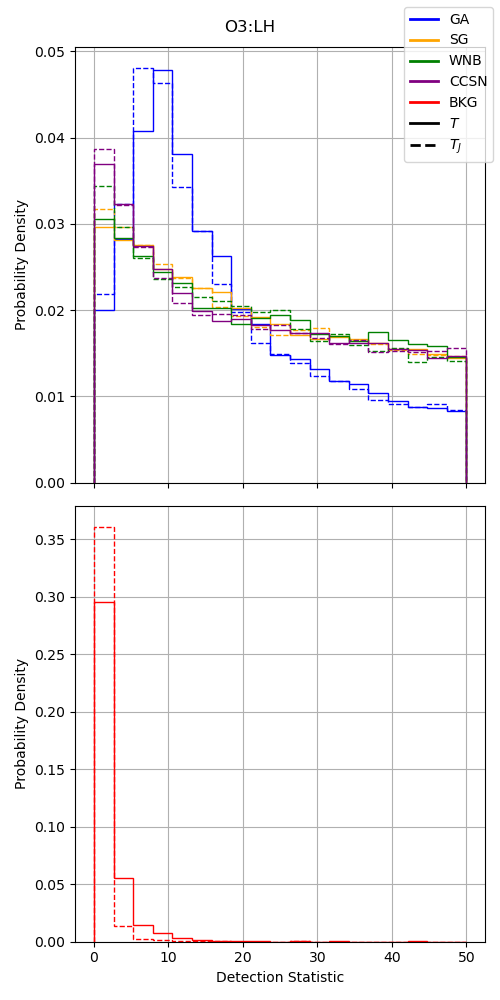}
    \includegraphics[width=0.45\textwidth]{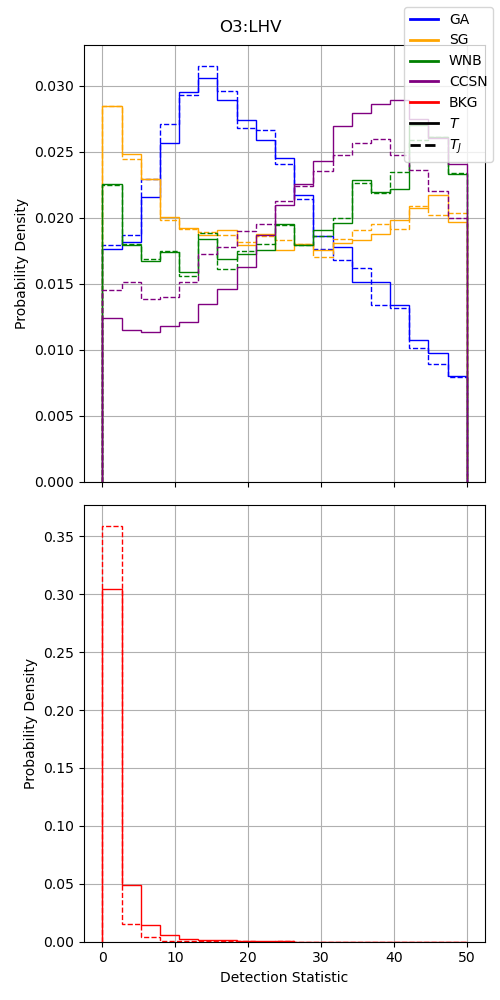}
    \caption{The left and right panels show the distributions for the LH and the LHV network. The top figures show the distributions for all types of simulations (each colour showing a particular type of signal, as labelled in the legends), and the bottom figures show the distributions for the background triggers. The solid histograms show the $T$ distributions, while the dashed histograms show the $T_J$ distributions. For the background distribution, all triggers lie below $T=50$. For signals, we show the distributions only up to detection statistic = 50, as these are the signals which are boosted in significance (see the text) and hence are relevant for the discussion in this section. But we remind the reader that the distributions for signals extend well beyond $T=50$ which are not being shown in this plot.}
    \label{fig:T-and-TJ-dist}
\end{figure*}
To explain the differences between the percentage improvements in sensitivity seen for different waveform types, we plot the distributions of $T$ and $T_J$ for all types of triggers considered in this work - \ac{GAs}, \ac{SGs}, \ac{WNBs}, \ac{CCSN} and the background triggers in Fig. \ref{fig:T-and-TJ-dist}. The background triggers for the LH analysis are spread over a total background time of $\sim 400$ years, while those for the LHV analysis are spread over a total background time of $\sim 300$ years (See \cite{Smith_2024} for exact numbers). Thus, the $T$ and $T_J$ thresholds at \ac{IFAR}s of 100 years and 10 years will be such that at most a few tens of background triggers will be above them. For both the LH and the LHV analyses, these loudest background triggers lie in the range $T \in (5,30)$. For a given signal type, the change in sensitivity at \ac{IFAR}s of 100 and 10 years depends on the extent to which these loudest background triggers are down-ranked and the extent to which the signals lying in the same approximate range are down-ranked. Although, both signals and background triggers are down-ranked, the $T_J$ values of the background triggers are considerably less than their $T$ values, while the $T_J$ values of the signals are only minimally less than their $T$ values. Thus, the signals lying in this $T$ range are boosted in significance, because the background triggers which were earlier obfuscating them have now been pushed to the left. On the other hand, signals which have $T$ values higher than that of the loudest background trigger are not affected by our re-weighting, because they are already assigned the highest significance in the search \footnote{This is usually done for all searches. If a signal has a detection statistic value above that of the loudest background trigger, the \ac{IFAR} is taken to be the total background time available, and thus this signal is assigned the maximum possible significance.} based on the $T$ statistic. Therefore, the signals which lie in the range $T \in (5,30)$ are the signals which are boosted by the re-weighting procedure proposed in this work. From Fig. \ref{fig:T-and-TJ-dist}, we see that among all the waveform types considered in this work, the $T$ distributions for the \ac{GAs} have the highest support in the (5,30) range. This explains why the \ac{GAs} have the highest percentage improvement in sensitivity among all waveform types.

\section{$h_{rss50}$ values of cWB-GMM and cWB-GMM-JSD searches}

In Fig. \ref{hrss50_o3ab}, we displayed the percentage changes in the $h_{rss50}$ values when cWB-GMM is replaced by cWB-GMM-JSD. Here, we tabulate the $h_{rss50}$ values in Table. \ref{tab:hrss50_table} for the interested reader.

\begin{table*}[p]
    \centering
    \setlength{\tabcolsep}{10pt}
    \begin{tabular}{c | c c | c c}
         \hline
         \hline

           & \multicolumn{4}{c}{$h_{rss50}$  $(\times10^{-22}$ $1 / \sqrt{Hz})$} \\
           \hline
          \multirow{2}{*}{Waveform} & \multicolumn{2}{c|}{LH network} & \multicolumn{2}{c}{LHV network}  \\ 
           & cWB-GMM & cWB-GMM-JSD & cWB-GMM & cWB-GMM-JSD \\
          \hline\hline
          
          {\bf Gaussian Pulse} & & & & \\
          $\tau=0.1$ms  & 2.26 & 2.1 & 3.9 & 3.63 \\
          $\tau=1.0$ms  & 1.67 & 1.33 & 2.32 & 2.0 \\
          $\tau=2.5$ms  & 2.6 & 1.87 & 3.15 & 2.62 \\
          $\tau=4.0$ms  & 3.89 & 2.75 & 4.3 & 3.71 \\
          \hline
          
          {\bf Sine-Gaussian} & & & & \\
          $f_0=70$Hz, Q=3 & 1.05 & 0.93 & 1.43 & 1.31 \\ 
          $f_0=70$Hz, Q=9 & 1.52 & 1.37 & 1.92 & 1.77 \\ 
          $f_0=70$Hz, Q=100 & 1.01 & 0.97 & 1.36 & 1.28 \\ 
          $f_0=100$Hz, Q=9 & 1.27 & 1.18 & 1.78 & 1.61 \\ 
          $f_0=153$Hz, Q=9 & 0.8 & 0.74 & 1.11 & 1.03 \\ 
          $f_0=235$Hz, Q=3 & 0.81 & 0.68 & 1.19 & 1.09 \\ 
          $f_0=235$Hz, Q=9 & 1.03 & 0.9 & 1.52 & 1.38 \\ 
          $f_0=235$Hz, Q=100 & 0.66 & 0.63 & 0.95 & 0.9 \\ 
          $f_0=361$Hz, Q=9 & 1.15 & 0.99 & 1.85 & 1.66 \\ 
          $f_0=554$Hz, Q=9 & 1.03 & 0.9 & 1.55 & 1.43 \\ 
          $f_0=849$Hz, Q=3 & 1.5 & 1.35 & 2.63 & 2.47 \\ 
          $f_0=849$Hz, Q=9 & 1.35 & 1.15 & 2.2 & 2.02 \\ 
          $f_0=849$Hz, Q=100 & 1.16 & 1.03 & 1.7 & 1.59 \\ 
          \hline

          {\bf White Noise Burst} & & & & \\
          $f_{low}=150$Hz  & 0.99 & 0.97 & 1.58 & 1.46 \\
          $f_{low}=300$Hz & 1.04 & 0.95 & 1.66 & 1.53 \\
          $f_{low}=700$Hz & 1.49 & 1.36 & 2.49 & 2.29 \\
          \hline

          {\bf Core-collapse Supernova} & & & & \\
          And s11 & 1.95 & 1.81 & 1.83 & 1.72 \\
          Mul L15 & 1.08 & 0.95 & 0.95 & 0.87 \\
          Pow he3.5 & 2.48 & 2.36 & 2.36 & 2.24 \\
          Rad s13 & 2.11 & 1.95 & 1.89 & 1.77 \\
          Rad s25 & 2.54 & 2.3 & 2.22 & 2.07 \\
          Rad s9 & 2.02 & 1.93 & 1.98 & 1.87 \\
          Pow s18 & 2.59 & 2.44 & 2.43 & 2.3 \\
          Kur SFHx & 1.13 & 1.02 & 0.96 & 0.88 \\
          Oco mesa20 & 3.31 & 3.01 & 2.94 & 2.7 \\
          Oco mesa20\_pert & 2.96 & 2.7 & 2.62 & 2.47 \\
          Abd A4O01.0 & 1.64 & 1.55 & 1.23 & 1.27 \\
          \hline

         \hline
         \hline
    \end{tabular}
    \caption{Table detailing the $h_{rss50}$ in units of $\times10^{-22}$ $1/\sqrt{Hz}$ achieved at an \ac{IFAR} $\geq$ 10 years for each injected waveform in O3 across for cWB-GMM and cWB-GMM-JSD.}
    \label{tab:hrss50_table}
\end{table*}

\bibliographystyle{apsrev4-1}
\bibliography{references}

\end{document}